\documentclass{appolb}
\usepackage{graphicx}

\begin{document}
\title{Diffractive Physics Program at the Electron-Ion Collider's (EIC) 2nd Detector
\thanks{Presented at ``Diffraction and Low-$x$ 2024'', Trabia (Palermo, Italy), September 8-14, 2024.}%
}
\author{Jihee Kim
\address{Department of Physics, Brookhaven National Laboratory, Upton, NY 11973, U.S.A.}
\\[3mm]
}
\maketitle
\begin{abstract}
The Electron-Ion Collider (EIC) will be a novel experimental facility to explore the properties of gluons in nucleons and nuclei, shedding light on their structure and  dynamics. The EIC community outlined the physics program of the EIC in a White Paper, and the demanding detector requirements and potential technologies to deploy at an EIC detector were published in a comprehensive Yellow Report. The general-purpose detector resulting from this efforts, ePIC,  is designed to perform a broad physics program. At the same time, the wider EIC community is strongly in favor of a second detector at the EIC. Having two general-purpose collider detectors to support the EIC science program, allows us to have cross-checks and control of systematic uncertainties for potential scientific discoveries. The second detector should feature complementary technologies where possible. It can also focus on specific measurements that are less well addressed by ePIC. The second interaction region provides potentially improved forward detector acceptance at low $p_{T}$ and a secondary beam focus that enables to enhance the  exclusive, tagging, and diffractive physics program. Hereby, I will present potential capabilities of the second detector and discuss studies related to its diffractive physics program.
\end{abstract}
  
\section{Introduction}
The Electron-Ion Collider (EIC) will be a pioneering experimental facility using Deep Inelastic Scattering (DIS) to investigate the inner structure of nucleons and nuclei. Its primary goal is to explore the properties of quarks and gluons within these particles. The EIC community has outlined the facility's physics program in a White Paper~\cite{Accardi:1498519}. To achieve this, the EIC will feature highly polarized electron beams colliding with ion beams ranging from protons to heavy nuclei. Additionally, the EIC will require a broad center-of-mass energy range and high luminosity to map the internal structure of nucleons and nuclei in the $x$ and $Q^{2}$ phase space. In terms of detector requirements, the EIC will need a comprehensive central detector with large rapidity coverage to track all final-state particles produced in the collisions. Specialized detectors will also be required in the forward (hadron-going) and backward (electron-going) directions, with optimal integration in the interaction region. The demanding detector requirements and potential technologies for the EIC were detailed in a comprehensive Yellow Report~\cite{ABDULKHALEK2022122447}.

The general-purpose detector resulting from this effort, ePIC (electron-Proton and Ion Collider), will be built at Interaction Point 6 (IP-6) and is designed to support the full EIC physics program. At the same time, the broader EIC community strongly advocates for a complementary second detector. The EIC is capable of supporting two Interaction Regions (IR-6 and IR-8), but currently only IP-6 will host a detector (ePIC). Having two general-purpose collider detectors will facilitate cross-checks of key results and data combination, helping to improve experimental systematics. Furthermore, the second detector may use different technologies to measure similar final states, with potentially different emphases, a concept known as complementarity in the design. It may also focus on distinct primary physics programs, optimizing sensitivity to the full EIC physics scope. Operating two interaction regions will require sharing the luminosity between the detectors at the same center-of-mass energy. The two interaction regions will offer distinct features: for example, the second interaction region will have different ``blind spots" due to the differing crossing angles. Additionally, it will provide larger forward detector acceptance at low transverse momentum for forward-scattered particles, enhancing the exclusive, tagging, and diffractive physics programs.

\begin{figure}[htb]
\centerline{%
\includegraphics[width=12.5cm]{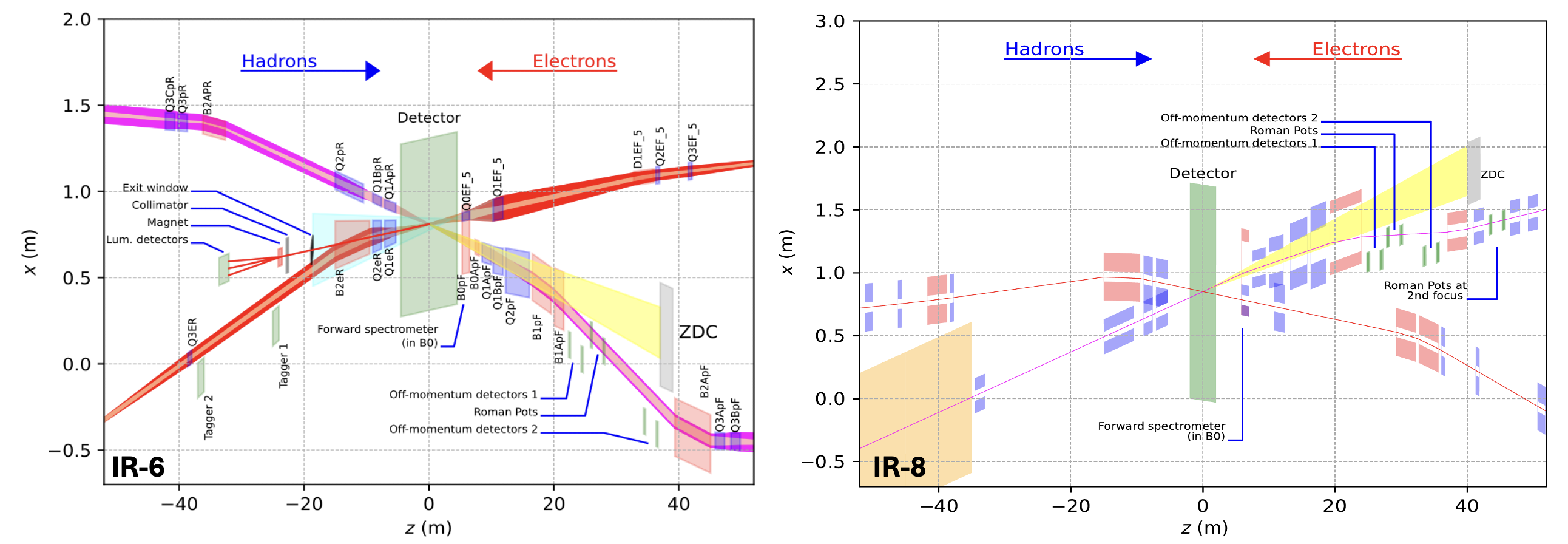}}
\caption{Left: Layout of the first interaction region (IR-6), featuring a 25 mrad crossing angle. Right: Layout of the second interaction region (IR-8), featuring a 35 mrad crossing angle, with lattice design and secondary beam focus position.}
\label{Fig:IRs}
\end{figure}

\section{Proposed IR-8 Layout}
The second EIC interaction region, shown in the right panel of Figure~\ref{Fig:IRs}, features a complementary design aimed at enhancing detection capabilities and optimizing, or potentially extending, the physics reach of the first interaction region. Similar to the first, the second interaction region is designed to fit within the existing experimental hall and includes beamline detector components for both the forward and backward regions. Due to the geometric constraints of the hall and tunnel, the optimal crossing angle for the second interaction region is proposed to be 35 mrad, larger than the 25 mrad angle of the first region. This increased angle results in different blind spots in pseudo-rapidity in the central detector. However, the proposed hadron beamline for the second interaction region introduces a secondary beam optics focus approximately 45 m downstream of the interaction point (IP-8), achieved by adding dipole and quadrupole magnets. The secondary focus creates a narrow beam profile similar to that at the interaction point, enabling detection of particles with very small changes in rigidity. This design offers complementarity to the first EIC detector, benefiting the exclusive, tagging, and diffractive physics programs. This complementarity, combined with the IR-6 configuration and detector layout, is crucial for the EIC in enhancing its scientific goals.

\section{Far-Forward Detectors}
For diffractive physics measurements, the EIC detector must provide broad acceptance for scattered charged and neutral particles, not only in the central detector but also in the far-forward region. Specialized detectors are essential in the interaction region to meet the needs of diffractive physics studies. The IR integrates multiple detectors along the beamline, including trackers and calorimeters to measure particle momenta and energy. The far-forward specialized detector system consists of the B0 spectrometer, the Off-Momentum Detector (OMD), the Zero-Degree Calorimeter (ZDC), and Roman Pots (RP). The primary purpose of the far-forward detectors is to tag scattered protons, ions, and nuclear fragments from diffractive events. By placing Roman Pot detectors around the secondary focus, these detectors can be positioned closer to the core of the beam, significantly improving forward detector acceptance for scattered protons or ions and nuclear fragments that would otherwise be undetectable due to their proximity to, or position within, the beam envelope.

\section{Examples of Physics Opportunities with the Secondary Focus in the Far-forward Region}
Specifically, the pre-conceptual design of IR-8 improves the acceptance of low transverse momentum for protons and light nuclei in exclusive reactions at very low $t$, thereby enhancing the potential for diffractive physics measurements at the EIC. This article introduces a few examples of these opportunities.

\subsection{Vetoing Efficiency}
One of the key measurements in the exclusive, tagging, and diffractive physics program at the EIC is the study of coherent diffractive vector meson production~\cite{Accardi:1498519}, which allows the spatial distribution of partons inside the nucleus to be probed. The recoiling target nucleus or its breakup products can be detected with specialized beamline detectors located meters away from the interaction point. Identifying coherent processes is critical because diffractive vector meson production includes both coherent (nucleus stays intact) and incoherent (nucleus breaks up) contributions. Tagging a coherent nucleus is experimentally challenging, especially for nuclei with larger A. Therefore, to distinguish between coherent and incoherent diffractive production, the experiment needs the capability to tag incoherent events, where the nucleus breaks up into fragments that are scattered into the far-forward direction, close to the hadron beamline, in the $e+$A program. Specialized detectors in the forward interaction region are designed to detect nuclear fragments, such as charged hadrons and neutral particles, traveling along the hadron beamline. The proposed IR-8 design is sufficient to suppress the incoherent contribution, serving as a complementary feature and providing a unique capability. It is important to note, however, that the current IR-8 design is in its very early stages, and further machine studies are required to determine how to operate the EIC with two interaction regions while ensuring stable and optimal performance.

\subsection{Lambda Spin Measurement}
A new $\Lambda$ polarization measurement via deep exclusive meson production, $e + p \rightarrow e' + K^{+} + \Lambda$~\cite{PhysRevC.109.055205}, has been proposed for the upcoming EIC. This measurement requires detecting the $\Lambda$ traveling close to the beam direction, with its decay particles measured by far-forward detectors, while $e'$ and $K^{+}$ can be detected within the central detector's acceptance. A low-energy configuration is more feasible since the decay vertex occurs before the B0 spectrometer, allowing the decay products to be reconstructed in it. However, at higher energies, $\Lambda$ decay may occur beyond the B0, but the measurement of neutral final-state particles becomes more feasible due to the larger far-forward acceptance. The second detector can be optimized with a baseline far-forward detector layout as a complementary measurement. However, this design presents challenges for the accelerator, particularly in providing a larger aperture to accommodate on/off-momentum protons and neutrons.

\subsection{Diffractive Longitudinal Structure Function}
The complementary design of IR-8 enhances far-forward acceptance and provides the opportunity to constrain the diffractive longitudinal structure function ($F_{L}^{D}$)~\cite{PhysRevD.105.074006}, which is sensitive to gluon content. Unlike the large rapidity gap method used by the H1 collaboration~\cite{2011}, the EIC IR design minimizes uncertainties in selecting diffractive events, thanks to the far-forward detectors. Specifically, the IR-8 design enables the tagging of protons with higher rigidity, providing complementary measurements and a unique capability to access very low $t$. The diffractive longitudinal structure function has two main contributions: Reggeon and Pomeron. Lower values of $x_{L} = p'/p_{\textrm{beam}}$ are more sensitive to the Reggeon contribution, while higher values of $x_{L}$ are more sensitive to the Pomeron contribution. By allowing the measurement of particles with low transverse momentum via the secondary focus, IR-8 enables the study of both Reggeon and Pomeron contributions at the same machine, potentially opening new opportunities to distinguish and study each contribution separately.

\subsection{Light Nuclei Spectator Tagging}
As another example, this approach can be extended to study light nuclei, such as He, to perform nucleon structure measurements that enhance our understanding of QCD. Through the polarized $^{3}$He beam configuration~\cite{doi:10.1142/S2010194516601022}, planned as part of the initial early EIC project, it will be possible to probe the neutron spin structure using $^{3}$He as an effective neutron target, a key topic addressed in the science program reviewed by the National Academy of Sciences (NAS)~\cite{NAP25171}. The first measurement of this at IP-6~\cite{FRISCIC2021136726} was carried out using double spectator tagging methods, measured by far-forward detectors. As a complementary measurement, this provides a cross-check for measuring the neutron spin structure. This study not only helps optimize detectors in the far-forward region at IR-8, but also provides valuable insights into the flavor dependence of Transverse Momentum Distributions and the role of gluons in nuclear binding, alongside the first detector. Additionally, it enables the study of proton and neutron structure at the same machine, with cross-checking and cross-calibration between the first and second EIC detectors and interaction regions.

\section{Summary}
We showed how the design of the second detector and interaction region (IR-8) enables cross-checking of results with the first project detector and cross-calibration, where the complementary design provides an opportunity to minimize systematic uncertainties. It can also focus on different primary physics programs to optimize overall sensitivity to the full EIC physics scope. Specifically, this complementary pre-conceptual design of IR-8 offers improved low transverse momentum acceptance for protons and light nuclei in exclusive reactions at very low $t$. While a more detailed and realistic design of IR-8 is needed to fully assess the benefits of the secondary focus, the current results highlight the potential to enhance the exclusive, tagging, and diffractive physics programs through improved far-forward acceptance with this design.

\bibliographystyle{unsrt}
\bibliography{diffraction_and_low_x_2024_Jihee_KIM}

\end{document}